\documentclass[aps, pra, twocolumn, 10pt, superscriptaddress, longbibliography, nofootinbib, floatfix]{revtex4-1}

\usepackage[T1]{fontenc}
\usepackage[utf8]{inputenc}
\usepackage[english]{babel}
\usepackage[colorlinks, bookmarks=false, citecolor=blue, linkcolor=red, urlcolor=blue]{hyperref}

\usepackage{amsmath}
\usepackage{amssymb}
\usepackage{amsfonts}
\usepackage{mathrsfs}
\usepackage{physics}
\usepackage{verbatim}
\usepackage{graphicx}
\usepackage{xcolor}
\usepackage{setspace}
\usepackage{braket}
\usepackage{bbold}
\usepackage{color}
\usepackage{bm}

\usepackage{nicefrac}
\usepackage{tipa}
\usepackage{wasysym}
\usepackage[abs]{overpic}
\usepackage{cancel}
\usepackage{booktabs}

\usepackage[export]{adjustbox}
\graphicspath{{Figures/}}

\makeatletter
\newcommand{\padA}{\affiliation{Dipartimento di Fisica e Astronomia “G. Galilei”, via Marzolo 8, I-35131, Padova, Italy.}}
\newcommand{\padB}{\affiliation{Padua Quantum Technologies Research Center, Universit{\`a} degli Studi di Padova, Italy.}}
\newcommand{\padC}{\affiliation{INFN, Sezioni di Padova \& Bari, Italy.}}
\newcommand{\padD}{\affiliation{Institute for Complex Quantum Systems, Ulm University, Albert-Einstein-Allee 11, 89069 Ulm, Germany.}}
\newcommand{\padE}{\affiliation{Dipartimento di Fisica, Università di Bari, I-70126 Bari, Italy.}}

\usepackage{orcidlink}
\newcommand{\orcidmarco}{\orcidlink{0009-0000-5194-3445}}
\newcommand{\orciddaniel}{\orcidlink{0000-0001-7658-3546}}
\newcommand{\orcidgiuseppe}{\orcidlink{0000-0002-7280-445X}}
\newcommand{\orcidilaria}{\orcidlink{0000-0002-3806-2034}}
\newcommand{\orcidsimone}{\orcidlink{0000-0002-8882-2169}}

\definecolor{smoothred}{HTML}{C5232F}
\definecolor{mygreen}{rgb}{0,0.5,0}
\definecolor{myblue}{rgb}{0,0,0.75}
\definecolor{mymagenta}{cmyk}{0,1,0,0.12}
\definecolor{applegreen}{rgb}{0, 0.5, 0.0}

\newcommand{\pbasis}{\mathrm{P}}
\newcommand{\Id}{\mathbb{1}}
\newcommand{\zpauli}{Z}
\newcommand{\qham}{\mathcal{H}}
\newcommand{\sign}[1]{\mathrm{sign}(#1)}
\newcommand{\nint}[1]{\lceil{#1}\rfloor}
\newcommand{\bitlength}{\ell}
\newcommand{\bonddim}{m}
\newcommand{\nqubits}{n}
\newcommand{\rsakey}{N}
\newcommand{\samplespower}{\gamma}
\newcommand{\asrpl}{\rho_{sr}}
\newcommand{\roundingop}{\kappa}
\newcommand{\lattice}{\Lambda}
\newcommand{\latticebasis}{\mathcal{B}}
\newcommand{\latticeparameter}{c}
\newcommand{\latticepermutation}{f}
\newcommand{\lllbasis}{\mathcal{D}}
\newcommand{\target}{\bm{t}}
\newcommand{\singleCVP}{(\lattice, \target)}
\newcommand{\latticepoint}{\bm{b}}
\newcommand{\lllvector}{\bm{d}}
\newcommand{\ncvp}{n_{\mathrm{\scriptscriptstyle CVP}}}
\newcommand{\nops}{\mathcal{T}}
\newcommand{\tfields}{h_{x}}
\newcommand{\tfieldoom}{\alpha}

\newcommand{\maxbitlength}{130}  % Max bit-length simulated
\newcommand{\maxnqubits}{256}  % Max number of qubits simulated
\newcommand{\ttnrecord}{100}  % Max bit-length factorized
\newcommand{\ttnnqubitsrecord}{64}  % Number of qubits for the max bit-length factorization
\newcommand{\ttngammarecord}{2}
\newcommand{\rsakeyrecord}{791339171587617359026543582309}
\newcommand{\pfactorrecord}{428949705601033}
\newcommand{\qfactorrecord}{1844829734709373}
\newcommand{\ncvprecord}{28980}
\newcommand{\sboundrecord}{137477}
\newcommand{\sbasisdimrecord}{12801}
\newcommand{\numsrpairsrecord}{11278}

\newcommand{\amp}{2.0 \pm 0.6}  % C1
\newcommand{\rate}{0.04 \pm 0.01}  % C2
\newcommand{\muu}{1.38 \pm 0.05}  %\mu
\newcommand{\rescparam}{8.3}  %\omega
\newcommand{\bestchipoly}{4.0 \pm 1.5}

\newcommand{\bootstrapnsamples}{50{,}000}

\newcommand{\bestxiexp}{13}

\newcommand{\bestgammaexp}{275.52}
\newcommand{\bestchiexp}{13.2 \pm 3.6}

\newcommand{\ampbonddim}{6.6 \pm 0.7}
\newcommand{\powerbonddim}{0.42 \pm 0.02}
\begin{document}

%%% Title %%%
%%% ***** %%%
\title{Integer Factorization via Tensor Network Schnorr’s Sieving}

%%% Authors %%%
%%% ******* %%%
\author{Marco Tesoro\orcidmarco}\padA \padB \padC
\author{Ilaria Siloi\orcidilaria}\padA \padB \padC
\author{Daniel Jaschke\orciddaniel}\padA \padB \padC \padD
\author{Giuseppe~Magnifico\orcidgiuseppe}\padA \padB \padC \padE
\author{Simone Montangero\orcidsimone}\padA \padB \padC

%%% Others %%%
%%% ****** %%%
% Comment for arXiv submission
% \date{\today}

%%% Abstract %%%
%%% ******** %%%
\begin{abstract}

\noindent Classical public-key cryptography standards rely on the Rivest-Shamir-Adleman (RSA) encryption protocol. The security of this protocol is based on the exponential computational complexity of the most efficient classical algorithms for factoring large semiprime numbers into their two prime components. Here, we address RSA factorization building on Schnorr’s mathematical framework where factorization translates into a combinatorial optimization problem.
We solve the optimization task via tensor network methods, a quantum-inspired classical numerical technique. This tensor network Schnorr’s sieving algorithm displays numerical evidence of polynomial scaling of resources with the bit-length of the semiprime. We factorize RSA numbers up to $\ttnrecord$ bits and assess how computational resources scale through numerical simulations up to $\maxbitlength$ bits, encoding the optimization problem in quantum systems with up to $\maxnqubits$ qubits.
Only the high-order polynomial scaling of the required resources limits the factorization of larger numbers. Although these results do not currently undermine the security of the present communication infrastructure, they strongly highlight the urgency of implementing post-quantum cryptography or quantum key distribution.

\end{abstract}
\maketitle

% ================================================================================
% Introduction                                                                   %
% ================================================================================

Among modern encryption schemes, the RSA protocol stands out for its resilience against cyber threats from conventional computers~\cite{rsa_1978}. The security of RSA encryption is rooted in the computational complexity of factoring any large semiprime $\rsakey$ in its two prime components $p$ and $q$, i.e., $\rsakey = p \cdot q$~\cite{galbraith2012_mathsOfpubkey,aggarwal2016_RSAequivFIP}.
State-of-the-art factoring algorithms for RSA numbers are sieving methods~\cite{crandall2006_PrimeNumbers, rabah2006_reviewFIP, boudot2022_review}, which rely on the observation that any odd number can be expressed as the difference of two squares $\rsakey = {\left(\frac{p + q}{2}\right)}^2 - {\left(\frac{p - q}{2}\right)}^2 = X^2 - Y^2$, where $X$ and $Y$ are integers. The search for prime factors then revolves around finding non-trivial solutions of the congruence relation $X^2 \equiv Y^2 \mod \rsakey$.
Sieving techniques differ in how potential values of $X$ and $Y$ are obtained. Then, it is efficient to compute the corresponding $p$ and $q$ and check if they factorize $\rsakey$~\cite{crandall2006_PrimeNumbers}.

The algorithmic complexity of sieving approaches has improved over the years through the application of sophisticated methods for constructing congruences of squares~\cite{pomerance2009_tale}. The most efficient known classical algorithm is the General Number Field Sieve (GNFS), which exhibits a sub-exponential algorithmic complexity ~\cite{stevenhagen2008_GNFS, kleinjung2010_RSA-768record} and has led to the successful factorization of the largest recorded RSA number ($829$ bits) in 2020~\cite{boudot2020_RSA-250record}. Shor’s quantum factorization algorithm constructs solutions to congruence relations~\cite{shor1997_quantumFIP} in polynomial time, marking a transformative advancement in the field. However, attacking state-of-the-art RSA remains out of reach for current near-term quantum devices~\cite{gidney2021_shor2048est}.
\begin{figure*}[t]
    \centering
    \includegraphics[width=0.9\textwidth]{Figures/schnorr_tensor_network_sketch}
    \caption{\emph{Sketch of the tensor network Schnorr’s sieving (TNSS) algorithm.}
    The factorization of the RSA number $\rsakey = p \cdot q$ into its prime components is mapped into an optimization problem on a lattice (\textbf{Lattice problem mapping}).
    This problem is initially approached by searching for approximate solutions to a series of $\rsakey$-related CVPs, each defined by a lattice-target pair $\singleCVP$ in $\nqubits + 1$ dimensions~\cite{schnorr2021_classicalSchnorr}. A representative of the CVP set for the simplest case $\nqubits = 1$ is depicted. The lattice basis $\latticebasis$ contains only one vector $\latticepoint_1 \in \mathbb{Z}^{2}$, which generates the infinite discrete set of lattice points $\latticepoint \in \lattice$ (black points). The blue point is the target point $\target \in \mathbb{Z}^2$ defining the CVP $\singleCVP$. A single approximate solution $\latticepoint^{cl} \in \lattice$ (orange point) to the CVP $\singleCVP$ is provided by Babai’s classical algorithm~\cite{babai1986_BabaiAlgorithm}. To improve over the classical solution, we look for additional lattice points potentially encoding sr-pairs in the eigenstates of a spin glass Hamiltonian describing an $\nqubits$-qubit system (\textbf{Spin glass mapping})~\cite{yan2022_quantumSchnorr}. The $2^\nqubits$ configurations correspond to the set of lattice points (light grey area) around Babai’s solution. The form of the Hamiltonian is defined in Eq.~\eqref{eqn:cvp_hamiltonian}.
    To compute the eigenstate of the spin glass Hamiltonian, we exploit TN methods (\textbf{Tree Tensor Network sieving}). By variationally minimizing the energy, we obtain a tree tensor network (TTN) state representing a superposition of low-energy eigenstates of $\qham$~\cite{silvi2019_anthology}. Finally, we sample a sufficient number of classical eigenstate configurations and extract their probabilities~\cite{ballarin2025_sampling}. 
    We discard all states violating the sr-pairs condition on the smoothness basis $\pbasis_2$ (gray lattice points) and collect the sr-pairs (green lattice points) from each CVP. These pairs are then combined to compute the two prime components $p$ and $q$ of the input RSA number $\rsakey$ in the final processing step.}
    \label{fig:algorithm_sketch}
\end{figure*}
In this work, we focus on Schnorr’s lattice sieving~\cite{schnorr1991_diophantine, schnorr2021_classicalSchnorr}, a classical algorithm that recasts RSA factorization as the search for optimal solutions to a set of NP-hard optimization problems on a lattice, known as Closest Vector Problems (CVPs)~\cite{emdeboas1981_cvpNPhardness, micciancio2001_hardnessCVP}. We improve Schnorr’s lattice sieving by introducing an approach based on tensor network (TN) methods~\cite{schollwoeck2011_dmrgmps, montangero2018_book}.
Following Yan et al.~\cite{yan2022_quantumSchnorr}, CVP solutions are encoded in the low-energy configurations of a spin glass Hamiltonian. We employ an optimal sampling strategy based on TN to efficiently access these configurations~\cite{ballarin2025_sampling}. Finally, we construct the two squares $X^2$ and $Y^2$ satisfying the congruence relation from the combination of these configurations.

The TN Schnorr’s sieving (TNSS) allows the factorization of RSA semiprimes up to over $\ttnrecord$ bits.
Moreover, we unveil how to effectively choose the hyperparameters of the algorithm for successful factorization with any key bit-length $\bitlength$~\cite{yan2022_quantumSchnorr, grebnev2023_pitfalls, khattar2023_commentGoogle, aboumrad2023_ionqSchnorr}. We provide a systematic numerical analysis of the resources needed by the TNSS algorithm up to $\maxbitlength$ bits using up to $\maxnqubits$ qubits, which scale polynomially with the key bit-length. Our findings, unless disproved by larger-scale simulations or mathematical proof, suggest that factoring RSA keys requires polynomial classical resources using the TNSS algorithm.

% --------------------------------------------------------------------------------
\subsection{The Tensor Network Schnorr’s Sieving factoring algorithm}
\label{subsec:tnss_algorithm}
% --------------------------------------------------------------------------------

We propose the TNSS factorization algorithm as sketched in Figure~\ref{fig:algorithm_sketch}. Our scheme builds on Schnorr’s idea of using lattice-based sieving to filter candidate congruence relations for factoring the semiprime $\rsakey$ (see Methods~\ref{subsec:methods_schnorr_sieving_details}). The first phase of Schnorr’s algorithm leads to the collection of a pool of candidates by searching for approximate solutions to the set of CVPs associated with $\rsakey$. In the second phase, further processing via linear algebra outputs the final prime components $p$ and $q$~\cite{schnorr1991_diophantine, coppersmith1993_linalgGF2, kleinjung2010_RSA-768record, schnorr2021_classicalSchnorr}. We leverage TNs to improve the efficiency of the collection phase, being exponentially time-consuming in the original formulation of the algorithm (see the discussion in Methods~\ref{subsec:methods_tnss_justification}).

\textit{Definitions.} Given a semiprime $\rsakey$ with bit-length $\bitlength = \lfloor{\log_2{\rsakey} + 1}\rfloor$, one defines a list of consecutive prime numbers, i.e., a \emph{factoring basis}, for decomposing  $\rsakey$: $\pbasis_1 = \left\{p_1, p_2, \dots, p_{\pi_1}\right\}$, where $\pi_1 = \nqubits$ is the size of $\pbasis_1$. Following the procedure outlined in Methods~\ref{subsec:methods_schnorr_sieving_details}, one constructs a set of CVPs associated with $\rsakey$. Each CVP considers an $\nqubits$-rank lattice $\lattice$ described by the lattice basis $\latticebasis = \{\latticepoint_1, \latticepoint_2, \dots, \latticepoint_{\nqubits}\}$, where $\latticepoint_j \in \mathbb{Z}^{\nqubits+1}$ for $j=1, \dots, \nqubits$. Further, one introduces a target vector $\target \in \mathbb{Z}^{\nqubits + 1}$ which does not belong to the lattice. The lattice-target pair $\singleCVP$ defines the CVP, where one searches for the closest lattice point $\latticepoint \in \lattice$ to the target $\target$~\cite{galbraith2012_mathsOfpubkey}.

The above construction associates to each lattice point a pair of integers (see Methods~\ref{subsec:methods_schnorr_sieving_details}). For a lattice point to be useful for factorization, each pair of integers must abide by the smoothness property: their ratio must not include prime factors larger than a chosen \emph{smoothness bound} $B_2 = p_{\pi_2}$. We define the \emph{smoothness basis}, i.e., a prime number basis $\pbasis_2$ of size $\pi_2$. Each pair, referred to as a smooth-relation (sr) pair, corresponds to a lattice point sufficiently close to the target $\target$~\cite{schnorr2021_classicalSchnorr}. Mathematically, each sr-pair represents a congruence relation between smooth integers modulo $\rsakey$.

\textit{Sieving, the collection phase.} Babai’s nearest plane algorithm~\cite{babai1986_BabaiAlgorithm, yang2023_improvedBabai} provides a single solution $\latticepoint^{cl} \in \Lambda $ to the CVP, representing an approximation to the closest lattice point to the target. For those CVPs where $\latticepoint^{cl}$ completely factorizes over the chosen smoothness basis $\pbasis_2$, Babai’s algorithm finds a sr-pair in polynomial time. To increase the pool of approximate solutions of a single CVP on $\Lambda$ and, thus, the number of sr-pairs, one constructs an optimization problem~\cite{yan2022_quantumSchnorr} that minimizes the Euclidean distance between the target $\target$ and lattice points around $\latticepoint^{cl}$. The corresponding all-to-all $\nqubits$-qubit Hamiltonian~\cite{yan2022_quantumSchnorr} reads:
\begin{equation}
    \label{eqn:cvp_hamiltonian}
    \qham
    =
    \sum_{k=1}^{\nqubits + 1}
    {\Bigg[
    t_k - b^{cl}_k -
    \sum_{j=1}^{\nqubits} \sign{\mu_j - c_j} d_{j, k} \zpauli_j
    \Bigg]}^2,
\end{equation}
where $\zpauli_j = \frac{1}{2}\big(\Id - \sigma^z_j\big)$, and $\sigma^z_j$ is the z-Pauli matrix. Each vector $\lllvector_j \in \mathbb{Z}^{\nqubits+1}$ for $j = 1, \dots, \nqubits$ is obtained by reducing the lattice basis $\latticebasis$ of $\lattice$ via LLL-reduction~\cite{lenstra1982_LLL}, while $ \{ \mu_j \}_{j=1}^{\nqubits} $ are the Gram-Schmidt coefficients from the orthogonalization of $\latticebasis$. Values of $\{c_j\}_{j=1}^{\nqubits}$ are derived by rounding each $\mu_j$ to the nearest integer: this is a fundamental approximation in Babai’s method~\cite{galbraith2012_mathsOfpubkey}. As $\qham$ is diagonal in the computational basis, each eigenstate corresponds to one of the $2^{\nqubits}$ possible roundings when constructing a new approximate solution around the classical one. Consequently, the eigenstates of $\qham$ encode additional lattice points and eventually lead to sr-pairs useful for factoring $\rsakey$. Moreover, the energy of each eigenstate gives the square of the distance between the associated lattice point and the target $\target$.

Heuristic arguments associate low-energy eigenstates with lattice points that may have a larger probability of being sr-pairs~\cite{schnorr1991_diophantine, schnorr2021_classicalSchnorr}. To improve the efficiency of the collection phase, we search for these states via TN optimization: approximating the wave function as a tree tensor network (TTN)~\cite{shi2006_ttnSimulation, montangero2018_book, silvi2019_anthology} and using a variational ground-state search algorithm, we find a TTN state with sufficient overlap with low-energy eigenstates (see Methods~\ref{subsec:methods_ttn_formulation_details} and~\ref{subsec:methods_ttn_sampling}). The lattice points are then extracted using an efficient sampling algorithm~\cite{ballarin2025_sampling}. Both algorithms are implemented in the Quantum TEA suite~\cite{quantumtea, qtealeaves}.

\textit{Processing phase.} Upon gathering at least $\pi_2 + 2$ independent sr-pairs, we apply a linear algebra step to pinpoint the subset of pairs to combine, as detailed in Methods~\ref{subsec:methods_schnorr_sieving_details}. From their combination, we obtain the two congruent squares $X^2$ and $Y^2$ for the derivation of $p$ and $q$ prime factors.
This post-processing step is efficiently implemented by leveraging index calculus and linear algebra techniques~\cite{bard2009_galois}.

% --------------------------------------------------------------------------------
\subsection{Addressing the curse of sr-pairs}
\label{subsec:n_qubits}
% --------------------------------------------------------------------------------

Collecting enough congruence relations between smooth numbers (sr-pairs) is the only requirement for successful factorization~\cite{pomerance2008_QuadraticSieve}. The number of qubits $\nqubits$ used to encode the RSA key, i.e., the lattice rank, and the smoothness bound $B_2 = p_{\pi_2}$ are tunable hyperparameters of the algorithm. Schnorr’s original prescription~\cite{schnorr2021_classicalSchnorr, yan2022_quantumSchnorr} suggests increasing both $\nqubits$ and $\pi_2$ sublinearly with the bit-length, $\nqubits= \pi_2 = \bitlength \mathbin{/} \log_2 \bitlength$. However, these choices prohibit the factorization of semiprimes larger than $\bitlength = 70$ bits in practice~\cite{ducas2022_schnorr, grebnev2023_pitfalls, khattar2023_commentGoogle, aboumrad2023_ionqSchnorr} due to the exponentially vanishing number of sr-pairs extracted from sublinear dimensional CVPs. For more details, we refer to the discussion in the Methods~\ref{subsec:methods_tnss_justification}.
\begin{figure*}[t]
    \begin{center}
        \begin{minipage}{0.48\linewidth}
            \begin{overpic}[width=1.0\columnwidth, unit=1mm]{AsrPL_vs_bitlength_rescaled}
                \put(0,55){a)}
            \end{overpic}
        \end{minipage}\hfill
        \begin{minipage}{0.48\linewidth}
            \begin{overpic}[width=1.0\columnwidth, unit=1mm]{number-of-qubits_scaling}
                \put(0,55){b)}
            \end{overpic}
        \end{minipage}\hfill
    \end{center}
    \begin{center}
        \caption{\emph{Scaling of the relevant resources in the TNSS algorithm.}
        \textbf{(a)}~The rescaled average number of sr-pairs per lattice (AsrPL) $\asrpl^{\text{eff}} = \asrpl \mathbin{/} \bitlength^\samplespower$ as a function of the effective bit-length $\bitlength_{\text{eff}} = \bitlength \mathbin{/} {\nqubits}^{1 \mathbin{/} {\omega}}$ of the input RSA number. The data are consistent with an exponential decay, see Eq.~\eqref{eqn:sr_pairs_density_scaling}, (dashed black line), characterized by fitted parameters $C_1 = \amp$, $C_2 = \rate$, and $\mu = \muu$. The grey-shaded area around the dashed black line accounts for the uncertainty in the fitted parameters and corresponds to a $3\sigma$ confidence interval. Each data point is obtained by averaging $\asrpl$ over $10$ random RSA keys. Each $\asrpl$ is estimated using $\ncvp = 50$ lattices.
        \textbf{(b)}~Scaling of the number of qubits~---~see Eq.~\eqref{eqn:n_qubits_scaling}~---~as a function of the bit-length $\bitlength$ and $\samplespower$.
        The hyperparameter $\samplespower$ tunes the number of samples extracted from each CVP Hamiltonian spectrum to achieve AsrPL $\asrpl = 1$. 
        The extrapolated number of qubits $\nqubits$ is required for factoring in a polynomial number of lattices within the $B_2 = 2 \nqubits \bitlength$ smoothness bound. Vertical dotted lines denote three RSA challenges~\cite{kaliski2005_encyclopedia}: RSA-250 ($829$ bits), marking the largest RSA key factorized to date~\cite{boudot2020_RSA-250record}, RSA-1024 and RSA-2048.}
        \label{fig:pairs_density_qubit_scaling}
    \end{center}
\end{figure*}
By leveraging TNs, we relax these conditions: to target more sr-pairs, we choose the smoothness bound to increase polynomially with the bit-length and the number of qubits, i.e., $\pi_2 = 2 \nqubits \bitlength$, and explore lattice sizes beyond Schnorr’s sublinear prescription (up to $\nqubits = \maxnqubits$ qubits). Further, a polynomial number of bit-strings $\order{\bitlength^\samplespower}$ is sampled from the TTN state, where $\samplespower$ is an integer hyperparameter controlling the number of low-energy configurations extracted from the single CVP. The TNSS algorithm successfully factorizes up to $\ttnrecord$-bit RSA keys using $\nqubits = \ttnnqubitsrecord$ qubits and $\samplespower = \ttngammarecord$. Details of the $\ttnrecord$-bit factorization are reported in Methods~\ref{subsec:methods_recorddetails}.

Choosing the size of the lattice according to the semiprime bit-length is crucial to ensure that enough sr-pairs are collected, once $\pi_2$ is set. To this aim, we consider bit-lengths up to $\bitlength = \maxbitlength$ bits, and, for each semiprime, we explore lattices of different sizes. For each $\nqubits$, we generate $\ncvp = 50$ CVPs, then sample a polynomial number of bit-strings from the TTN state. Figure~\ref{fig:pairs_density_qubit_scaling}a) shows that the \emph{average number of sr-pair per lattice} (AsrPL) $\asrpl$ obeys:
\begin{equation}
    \label{eqn:sr_pairs_density_scaling}
    \asrpl(\bitlength, \nqubits, \samplespower)
    \sim
    C_1 {\bitlength}^\samplespower
    \exp\left\{
    -C_2 {\left[\dfrac{\bitlength}{{\nqubits}^{1 \mathbin{/} \omega}}\right]}^\mu
    \right\},
\end{equation}
where $ C_1, C_2, \mu \in \mathbb{R}$ are fitting parameters and $\omega \approx \rescparam$.
Each data point in Figure~\ref{fig:pairs_density_qubit_scaling}a) is the average over $10$ semiprimes randomly drawn from an RSA key generator~\cite{rsa_generator}. The value of $\omega$ and the validity of the statistical model have been verified through a bootstrapped chi-squared test and a collapse quality score analysis.
To discern between a polynomial and a power-law scaling, we performed additional checks to confirm the distinguishability of the two models, as detailed in Methods~\ref{subsec:methods_chisquare}.
We emphasize that the scaling of the number of qubits in Eq.~\eqref{eqn:n_qubits_scaling} obeys a power-law function of the input bit-length $\bitlength$, and not a high-order polynomial, making the distinction between the two models statistically resolvable within the simulated range.

The scaling of the AsrPL shows again an exponential decay. This behavior is consistent with theoretical expectations from number theory, specifically the Dickman function, which describes the distribution of smooth numbers over integers~\cite{crandall2006_PrimeNumbers, aboumrad2023_ionqSchnorr}.
However, as a result of the chosen hyperparameters, Eq.~\eqref{eqn:sr_pairs_density_scaling} highlights a dependence on the effective bit-length, $\bitlength \mathbin{/} {\nqubits}^{1 \mathbin{/} \omega}$. Finally, the scaling for the number of qubits $\nqubits$ reads:
\begin{equation}
    \label{eqn:n_qubits_scaling}
    \nqubits(\bitlength, \asrpl, \samplespower)
    \sim
    {\Bigg[
    \dfrac
    {C_2 {\bitlength}^\mu}
    {\log{C_1} - \log{\asrpl} + \samplespower \log{\bitlength}}
    \Bigg]}^{\dfrac{\omega}{\mu}},
\end{equation}
where the size of the Hilbert space constraints the number of sampled bit-strings to the order $\samplespower \leq {\nqubits \log2} \mathbin{/} {\log \bitlength}$.

Figure~\ref{fig:pairs_density_qubit_scaling}b) shows the polynomial scaling of the number of qubits with the key size when sampling a polynomial number of bit-strings as dictated by Eq.~\eqref{eqn:n_qubits_scaling}, assuming to find at least one sr-pair for each lattice, i.e., $\asrpl = 1$. The more samples we collect, the smaller the required lattice size. The presence of at least one sr-pair per lattice allows targeting RSA keys within a polynomial number $\ncvp = \order{2 \nqubits \bitlength}$ of CVPs. This guarantees that at least one non-trivial combination of sr-pairs forms a congruence of squares modulo $\rsakey$, see Methods~\ref{subsec:methods_schnorr_sieving_details}.
Although the resources needed to factorize state-of-the-art RSA keys are still beyond current capabilities, these numerical findings indicate that the resources required by the TNSS scale polynomially with the input bit-length. In Eq.~\eqref{eqn:tnss_time_complexity} of Section~\ref{subsec:resources}, we provide the scaling of the number of operations for the entire factoring procedure that follows from Eq~\eqref{eqn:n_qubits_scaling}.

% --------------------------------------------------------------------------------
\subsection{Sieving via tree tensor networks}
\label{subsec:sampling_performance}
% --------------------------------------------------------------------------------

\begin{figure*}[t]
    \begin{center}
        \begin{minipage}{0.21\textwidth}
            \begin{overpic}[width=0.95\linewidth, unit=1mm]{ttn_no_sr_pairs_cloud_single_lattice.png}
                \put(0, 40){a)}
            \end{overpic}
        \end{minipage}
        \begin{minipage}{0.21\textwidth}
            \begin{overpic}[width=0.95\linewidth, unit=1mm]{ttn_sr_pairs_cloud_single_lattice.png}
                \put(0, 40){b)}
            \end{overpic}
        \end{minipage}
        \begin{minipage}{0.22\textwidth}
            \begin{overpic}[width=0.95\linewidth, unit=1mm]{ttn_pairs_sampling_prob_single_lattice.png}
                \put(0, 40.4){c)}
            \end{overpic}
        \end{minipage}
        \begin{minipage}{0.20\textwidth}
            \begin{overpic}[width=0.95\linewidth, unit=1mm]{max_bond_dimension_scaling.png}
                \put(0, 39.4){d)}
            \end{overpic}
        \end{minipage}
    \end{center}
    \begin{center}
        \caption{\emph{Sampling smooth-relation pairs from a single lattice via tree tensor networks.}
        \textbf{(a)--(b)}~Lattice points sampled from a CVP Hamiltonian using the efficient TTN OPES method~\cite{ballarin2025_sampling} with bond dimension $\bonddim = 8$. The input RSA key has a bit-length of $\bitlength = 70$ (dashed black horizontal line), and the CVP problem is characterized by $\nqubits = 32$ qubits. The number of samples is approximately $70^3$. No-sr pairs are denoted by gray dots in \textbf{(a)}, while green dots represent sr-pairs in \textbf{(b)}. The x-axis indicates the square root of the energy of the extracted eigenstate, i.e., the Euclidean distance to the target $\Vert \target - \latticepoint_j \Vert$ of the pair $j$ identified with the lattice point $\latticepoint_j$. The y-axis represents the size of the sampled pair $\bitlength_j$ measured in bits. The measured average number of sr-pairs per lattice (AsrPL) on the chosen smoothness bound $\pi_2 = 2 \nqubits \bitlength = 4480$ is $\asrpl = 432$. For both scatter plots, we also illustrate the probability distribution of the pairs with respect to their distance from the target (top panel) and their bit-length (right panel).
        \textbf{(c)}~The probability of sampling the first $400$ pairs, ordered from most to least probable for both sr (green bars) and no-sr (gray bars).
        \textbf{(d)}~The maximum values of the TTN bond dimension $\bonddim$ as a function of the number of qubits $\nqubits$. The data show the sublinear scaling with the number of qubits $\bonddim(\nqubits) \sim A {\nqubits}^\zeta$ with fitted parameters $A = \ampbonddim$ and $\zeta = \powerbonddim \approx 2 \mathbin{/} 5$ (dashed dark blue curve).}
        \label{fig:ttn_sampling_with_probability}
    \end{center}
\end{figure*}

The core of the TNSS is the OPES algorithm introduced in Ref.~\cite{ballarin2025_sampling}. This algorithm enables efficient sampling of bit-strings in the tail of a distribution from a TTN state, here representing a broad superposition of low-energy eigenstates of the CVP Hamiltonian in Eq.~\eqref{eqn:cvp_hamiltonian}, see Methods~\ref{subsec:methods_ttn_sampling}. The key feature of the OPES algorithm is its ability to avoid re-sampling the same bit-string regardless of its probability within the TTN state. This allows us to extract bit-strings with very low probabilities while preventing multiple samples of those with high probabilities, then optimizing the sieving of sr-pairs.

The effectiveness of this approach is demonstrated in Figure~\ref{fig:ttn_sampling_with_probability}, which shows the results of the sieving procedure on a single lattice with $\nqubits = 32$ for a semiprime with $\bitlength = 70$.
The TN sieving efficiency also depends on the bond dimension $\bonddim$ (see Methods~\ref{subsec:methods_ttn_formulation_details}), the size of the indices of each tensor in the network~\cite{montangero2018_book,silvi2019_anthology}.
We set $\bonddim = 8$ and sample approximately $\bitlength^3$ bit-strings. The scatter plots in Figure~\ref{fig:ttn_sampling_with_probability}a) and b) show both sr-pairs (green dots) and no-sr pairs (gray dots) in terms of their distance from the target point $\Vert \target - \latticepoint_j \Vert$, i.e., the square root of the bit-string energy ($x$-axis), and the size of the pair $j$ in bits $\bitlength_j$ ($y$-axis), see Methods~\ref{subsec:methods_ttn_sampling}.
All pairs accumulate in a low-energy region (lattice points close to the target) and cluster around the input bit-length $\bitlength = 70$~\cite{schnorr2021_classicalSchnorr, aboumrad2023_ionqSchnorr}, as shown by the probability distributions of green and gray dots in the panels on the right and at the top of each scatter plot.

In Figure~\ref{fig:ttn_sampling_with_probability}c), we plot the probability of sampling a bit-string encoding one of the sieved pairs. For both sr and no-sr cases, we consider the first $432$ extracted pairs with the largest probability. As clearly shown, the pairs with higher probabilities (gray bars) are those not useful for factoring. However, our sampling strategy excludes the possibility of measuring the same state more than once, thus efficiently sampling sr-pairs with lower probabilities (green bars). Notably, bond dimension $\bonddim = 8$ is sufficient to obtain $\asrpl = 432$ independent sr-pairs despite their small probabilities.

Lastly, Figure~\ref{fig:ttn_sampling_with_probability}d) displays the maximum bond dimension $\bonddim$ used to represent the variationally optimized many-body quantum state, i.e., the solution to the CVP Hamiltonian in Eq.~\eqref{eqn:cvp_hamiltonian}, as a function of the number of qubits $\nqubits$. The fit curve (dashed blue curve) indicates a sublinear growth of the bond dimension with $\nqubits$. These plots demonstrate the scalability of the TTN approach in solving the classical optimization task of searching for sr-pairs within the spectrum of CVP Hamiltonians, that are characterized by long-range glassy interactions.

% --------------------------------------------------------------------------------
\subsection{The Tensor Network Schnorr’s Sieving computational cost}
\label{subsec:resources}
% --------------------------------------------------------------------------------

We determine the scaling of the number of operations in the tensor network approach to Schnorr’s sieving. To this end, we assume $\asrpl = 1$, resulting in approximately $\ncvp \sim 2 \nqubits \bitlength$ needed for the successful factorization of RSA keys with bit-length $\bitlength$ using $\nqubits$ qubits.

The leading complexity in Schnorr’s original factoring method arises from Babai’s algorithm~\cite{galbraith2012_mathsOfpubkey, yan2022_quantumSchnorr}. In TNSS, we perform this step for each CVP to obtain the approximate closest vector $\latticepoint^{cl}$ to the target. The lattice point $\latticepoint^{cl}$ is then used to define the Hamiltonian in Eq.~\eqref{eqn:cvp_hamiltonian}. The number of operations $\nops_1$ to execute Babai’s algorithm scales as
\begin{equation}
    \nops_1
    =
    \order{\nqubits^7 \ \bitlength \ \log^3_{10}\nqubits} .
    \label{eqn:babai_time_complexity}
\end{equation}

The TN sieving step involves the variational energy optimization of the initial TTN state with bond dimension $\bonddim$, followed by the OPES sampling of lattice points for each CVP. If we sample $\bitlength^\samplespower$ pairs per CVP, the TN-based part $\nops_2$ accounts for the following number of operations (see Methods~\ref{subsec:methods_ttn_sampling} for details)
\begin{equation}
    \nops_2
    =
    \order{\nqubits^4 \ \bitlength \ \bonddim^4
    +
    \nqubits^2 \ \bitlength^{\samplespower + 1} \ \bonddim^4} .
    \label{eqn:ttn_time_complexity}
\end{equation}

Once the pairs are sampled, we construct the corresponding integers by computing the lattice point components on the basis $\latticebasis$ (see Methods~\ref{subsec:methods_schnorr_sieving_details}). Sr-pairs are identified by decomposing these integers on the smoothness basis $\pbasis_2$.
The worst-case scenario corresponds to applying a trial division strategy. In this case, the total complexity $\nops_3$ is bounded by
\begin{equation}
    \nops_3
    =
    \order{
    \nqubits^4 \ \bitlength
    +
    \nqubits^3 \ \bitlength^{\samplespower + 1}
    +
    \nqubits^2 \ \bitlength^{\samplespower + 2}
    }.
    \label{eqn:pairs_formation_time_complexity}
\end{equation}
Given that the bond dimension $\bonddim$ grows as $\nqubits^{2\mathbin{/}5}$, according to Figure~\ref{fig:ttn_sampling_with_probability}d), the time complexity $\nops_3$ in Eq.~\eqref{eqn:pairs_formation_time_complexity} is always sub-leading compared to the TN sieving step (see scaling in Eq.~\eqref{eqn:ttn_time_complexity}).
Overall, the number of operations $\nops = \nops_1 + \nops_2$ scales as
\begin{equation}
    \nops
    = 
    \order{
    \nqubits^7 \ \bitlength \ \log^3_{10}\nqubits
    +
    \nqubits^4 \ \bitlength \ \bonddim^4
    +
    \nqubits^2 \ \bitlength^{\samplespower + 1} \ \bonddim^4
    }.
    \label{eqn:tnss_time_complexity}
\end{equation}
The balance between Babai’s complexity $\nops_1$ and TTN complexity $\nops_2$ in Eq.~\eqref{eqn:tnss_time_complexity} depends on the input bit-length $\bitlength$ and the number of eigenstates sampled to obtain a sufficient set of sr-pairs, as determined by the hyperparameter $\samplespower$. As illustrated in Figure~\ref{fig:time_complexity}, Babai’s algorithm dominates up to $\samplespower < 8 $ for relevant key lengths of approximately $\bitlength \sim 10^{3}$.

In conclusion, we achieve an asymptotic polynomial scaling of the number of operations $\nops \lesssim \order{\bitlength^{59} \log^3_{10}\bitlength}$ for $\samplespower < 25$ based on the scaling for the number of qubits in Eq.~\eqref{eqn:n_qubits_scaling}. Furthermore, the memory requirements for the TN sieving process remain polynomial. The scaling of the bond dimension $\bonddim$ with the number of qubits shown in Figure~\ref{fig:ttn_sampling_with_probability}d) ensures that the wave function description of CVP solutions does not scale exponentially with $\nqubits$. For a single CVP, the memory needed for the TTN state and the Hamiltonian operator scales at most as $\order{\nqubits \bonddim^3} + \order{\nqubits^3 \bonddim^2} \lesssim \order{\bitlength^{32}}$ bits, while the sampling routine stores up to $\order{\bitlength^\samplespower}$ $\nqubits$-bit strings.

\begin{figure}[t]
    \centering
    \begin{overpic}[width=0.95 \columnwidth,unit=1mm]{Figures/tnss_time-complexity}
    \end{overpic}
    \caption{\emph{Scaling of the number of operations for the TNSS algorithm.}
    Scaling of the number of operations as a function of the input size $\bitlength$ for the TNSS algorithm derived from the predicted scaling of the number of qubits in Eq.~\eqref{eqn:n_qubits_scaling} and Eq.~\eqref{eqn:tnss_time_complexity}. We set the average number of sr-pairs per lattice (AsrPL) in Eq.~\eqref{eqn:n_qubits_scaling} to $\asrpl = 1$. Solid curves represent Babai’s contribution $\nops_1$ (first term in Eq.~\eqref{eqn:tnss_time_complexity}), dashed curves the TTN sieving term $\nops_2$ in Eq.~\eqref{eqn:tnss_time_complexity}, for various values of the hyperparameter $\samplespower$. Vertical dotted lines denote three RSA challenges~\cite{kaliski2005_encyclopedia}: RSA-250 (829 bits) marking the largest RSA key factorized to date~\cite{boudot2020_RSA-250record}, RSA-1024, and RSA-2048. The leading contribution in the algorithm varies with the number of sampled pairs: below $\samplespower = 8$, Babai’s algorithm prevails over the computation required for sampling lattice points in the range of currently relevant bit-lengths of $\bitlength \sim 10^3$. Within this range, the TN term $\nops_2$ in Eq.~\eqref{eqn:tnss_time_complexity} becomes the leading one for $\samplespower > 8$.}
    \label{fig:time_complexity}
\end{figure}

When compared to the GNFS~\cite{lenstra1993_GNFS, stevenhagen2008_GNFS}, the TNSS scaling of the number of operations is larger within the RSA key bit-length range relevant to current security protocols. Consequently, the GNFS remains the leading classical algorithm for factoring RSA. However, 
in the limit of large $\bitlength$, the TNSS approach becomes more efficient due to its predicted polynomial complexity.
%%% ******* %%%
%%% ******* %%%

% ================================================================================
\subsection{Conclusions}
\label{sec:discussion}
% ================================================================================

We have introduced a classical algorithm for RSA key factorization based on tensor network sieving applied to Schnorr’s lattice method. We achieve successful factorization by fine-tuning the hyperparameters of the TNSS algorithm, e.g., the number of qubits $\nqubits$, the smoothness basis size $\pi_2$, and the number of extracted eigenstates via $\samplespower$. The algorithm effectiveness spawns from the efficient TTN optimization and the OPES sampling.

The TNSS algorithm successfully factorizes randomly generated RSA numbers up to $\ttnrecord$ bits, validating the full pipeline of lattice generation, sieving, and smooth-related pairs post-processing. To the best of our knowledge, this is the largest RSA number factorized using Schnorr’s sieving method.

The algorithm is quantum inspired as it leverages concepts and methods from quantum algorithms. However, due to the tensor network approach and the limited amount of quantum correlations generated during the process, it can be efficiently run on classical computers.
We foresee that employing a universal scalable quantum computer to execute the same algorithm would not be beneficial: due to the distribution shown in \mbox{Figure~\ref{fig:ttn_sampling_with_probability}}, the sampling of the classical configurations on the quantum computer’s wave function would be highly inefficient, necessitating the development and exploration of alternative strategies.

We emphasize that the TNSS algorithm’s limitation in factoring larger keys lies in the high-order polynomial scaling of resources. We have identified Babai’s algorithm and the TN sampling of sr-pairs as the primary contributors to the overall scaling. However, TN efficient OPES sampling has emerged as crucial for acquiring sr-pairs useful for factoring $\rsakey$. This is mainly due to its ability to avoid re-sampling.

The numerical simulations have been executed on a single-node machine and will benefit from massive parallelization on multi-node architectures. Indeed, the algorithm features several highly parallelizable steps, such as the TN sampling and simultaneous CVP sieving: leveraging high-performance computing, parallelization, and code optimization could significantly improve its performance.

In conclusion, we have numerically investigated the scaling of classical resources required by the TNSS algorithm and observed behavior consistent with polynomial scaling over the studied range of bit-lengths $\bitlength \in \left[50, \maxbitlength\right]$. These findings suggest that tensor networks could provide a novel computational framework for addressing lattice-based cryptographic problems. While our conclusions are based on numerical evidence, and not on formal complexity-theoretical proofs, the observed trends indicate that factoring RSA integers using this approach may be achievable with polynomial classical resources~---~pending further validation at larger input sizes or through analytical investigation.
%%% ********** %%%
%%% ********** %%%

%%% Acknowledgments %%%
%%% *************** %%%
% ================================================================================
\section*{Acknowledgments}\label{sec:acknowledgments}
% ================================================================================
We thank M. Ballarin, M. Di Liberto, M. Rigobello, P. Silvi and E. Altelarrea-Ferr{\'e} for useful discussions.
We thank R. Fazio, T. Osborne, M. Montangero, S. Pascazio, F. Pollmann, N. Schuch, M. Stoudenmire, F. Verstraete, and P. Zoller for feedback on our work and on the manuscript.
We thank L. Zangrando for technical support on the computational side.

This work is partially supported by the WCRI~---~Quantum Computing and Simulation Center (QCSC) of Padova University, the Departments of Excellence grant 2023-2027 Quantum Frontiers and QuaSiModO, the Italian PRIN2022 project TANQU, the INFN IS-QUANTUM, the QuantERA projects TNISQ and QuantHEP, the European Union Horizon 2020 research and innovation program (Quantum Flagship~---~PASQuanS2, Euryqa), the European Union~---~NextGenerationEU project CN00000013~---~Italian Research Center on HPC, Big Data and Quantum Computing, the program PON “Ricerca e Innovazione” 2014-2020 project 19-G-12542-2, the University of Bari through the 2023-UNBACLE-0244025 grant, and the German Federal Ministry of Education and Research (BMBF) via the project \mbox{QRydDemo}.
We thank EuroHPC for the project \mbox{101194322~---~QEC4QEA}.
We acknowledge computational resources from Cloud Veneto, CINECA, BwUniCluster, and the University of Padova Strategic Research Infrastructure Grant 2017: “CAPRI: Calcolo ad Alte Prestazioni per la Ricerca e l’Innovazione”.
%%% *************** %%%
%%% *************** %%%

%%% ******* %%%
%%% Methods %%%
%%% ******* %%%
% \appendix*

\clearpage
% ================================================================================
\section*{Supplementary Material}
\label{sec:methods}
\setcounter{subsection}{0}
% ================================================================================
The appendix includes a detailed introduction to Schnorr’s lattice sieving in Section~\ref{subsec:methods_schnorr_sieving_details}, the description of our tensor network formulation in Section~\ref{subsec:methods_ttn_formulation_details} and of the sampling formalism in OPES in Section~\ref{subsec:methods_ttn_sampling}. In Section~\ref{subsec:methods_tnss_justification} we validate and compare our approach to the original Schnorr’s formulations and Yan \textit{et al.} extension \cite{yan2022_quantumSchnorr}. In Section~\ref{subsec:methods_recorddetails} we provide the prime factors of our record RSA-key factorization with TNSS algorithm. Finally, in Section~\ref{subsec:methods_chisquare} we describe the fitting procedure and how to choose the functions to rescale the bit-lengths in Figure~\ref{fig:pairs_density_qubit_scaling}a).

% --------------------------------------------------------------------------------
\subsection{Schnorr’s lattice-based factoring algorithm}
\label{subsec:methods_schnorr_sieving_details}
% --------------------------------------------------------------------------------

In this Section, we describe in detail Schnorr’s original procedure~\cite{schnorr1991_diophantine, schnorr2021_classicalSchnorr} for decomposing large semiprimes, as well as the quantum variant proposed in Ref.~\cite{yan2022_quantumSchnorr}.

\textit{Factoring by congruence of squares.} Schnorr’s factoring algorithm is rooted in the congruence of squares method~\cite{crandall2006_PrimeNumbers, galbraith2012_mathsOfpubkey}. Given a semiprime number $\rsakey = p \cdot q$, the algorithm determines the prime factors $p$ and $q$ by constructing the congruence $X^2 \equiv Y^2 \mod \rsakey$, with $X \not\equiv \pm Y \mod \rsakey$ being integers. Then, the algorithm efficiently computes $p = \gcd(X + Y, \rsakey)$ and $q = \gcd(X - Y, \rsakey)$ using the Euclidean algorithm~\cite{wagstaff2013_joy}. Many classical and quantum factorization techniques~\cite{pomerance2009_tale, pomerance2008_QuadraticSieve, boudot2022_review, lenstra1993_GNFS}, including Shor’s algorithm~\cite{shor1997_quantumFIP}, employ this strategy, differing in how they construct the above congruence. In Schnorr’s integer factorization, this process comprises two stages: an initial \emph{collection phase} followed by subsequent \emph{processing phase}.

\textit{Collection phase: smooth numbers.} The objective of the collection phase is to accumulate a sufficient number of so-called \emph{smooth-relation pairs} (sr-pairs). These pairs are defined based on smooth numbers, a pivotal concept in number theory and essential for factoring tasks~\cite{crandall2006_PrimeNumbers, pomerance2009_tale}. We introduce the set of the first $\pi$ consecutive prime numbers
\begin{equation}
    \pbasis = \left\{p_1, p_2, \dots, p_{\pi}\right\}
    \label{eqn:methods_factoring_basis}
\end{equation}
and call $B = p_{\pi}$ the smoothness bound. An integer is said to be smooth with respect to this bound if all its prime factors are less than or equal to $p_{\pi}$. Every $B$-smooth integer $u$ can be represented on the prime basis $\pbasis$ as
\begin{equation}
    u = \prod_{p_j \in \pbasis} {p_j}^{e_j}.
    \label{eqn:methods_smooth_integer}
\end{equation}
The exponents $\left\{e_j\right\}_{j=1}^{\pi} \in \mathbb{Z}^{\pi} $ define the multiplicities in the prime decomposition of $u$. Negative $B$-smooth integers are equivalently represented on the prime basis $\pbasis$ by including an extra element $p_0 = -1$ in Eq.~\eqref{eqn:methods_factoring_basis}. Thus, negative numbers are distinguished by the exponent $e_0 = 1$, while positive numbers have $e_0 = 0$.
As the prime decomposition is unique, we can represent a smooth number by the corresponding multiplicity vector $u \sim \bm{e} = \left(e_0, e_1, \cdots, e_{\pi}\right)$.
We define an integer pair $\left(u, v\right)$ as a \emph{smooth pair} if both $u$ and $v$ are $B$-smooth. If the integer $u - v\rsakey$ also turns out to be $B$-smooth, the triplet $\left(u, v, u - v\rsakey\right)$ is mathematically referred to as a \emph{fac-relation}~\cite{schnorr1991_diophantine}. We denote $\left(u, u - v\rsakey\right)$ as a smooth-relation (sr) pair. By construction, the integers forming an sr-pair satisfy $u \equiv u - v\rsakey \mod \rsakey $, leading to the following congruence
\begin{equation}
    \dfrac{u - v\rsakey}{u} \equiv 1 \mod \rsakey.
    \label{eqn:methods_sr_pair_definition}
\end{equation}
Additionally, we can represent the ratio of the $B$-smooth integers $u - v\rsakey$ and $u$ in Eq.~\eqref{eqn:methods_sr_pair_definition} using the corresponding multiplicity vectors on the common prime basis $\pbasis$:
\begin{align}
    \dfrac{u - v\rsakey}{u}
    &=
    \dfrac{\prod_{p_j \in \pbasis}{p_j}^{e^{\prime}_j}}{\prod_{p_j \in \pbasis} {p_j}^{e_j}} \nonumber \\
    &= \prod_{p_j \in \pbasis} {p_j}^{\left(e^{\prime}_j - e_j\right)} = \prod_{p_j \in \pbasis} {p_j}^{\tilde{e}_j}
    \label{eqn:methods_sr_pair_integer_definition}\\
    &\sim \bm{\tilde{e}} = \left(\tilde{e}_0, \tilde{e}_1, \dots, \tilde{e}_{\pi}\right). \nonumber
\end{align}
The most computationally demanding part of the factoring procedure consists of collecting a sufficient number $D$ of sr-pairs $\left\{\left(u, u - v\rsakey\right)\right\}_{r=1}^D$, where $D$ depends on the smoothness bound $B$~\cite{pomerance2008_QuadraticSieve}.

\textit{Processing phase via linear algebra in $\mathbb{GF}(2)$.} Schnorr’s approach involves leveraging the simple multiplicative structure of sr-pairs $\left(u - v\rsakey\right) \mathbin{/} u$ to construct two congruent squares $X^2$ and $Y^2$, thereby factorizing $\rsakey$. Once $D$ ratios of the form defined in Eq.~\eqref{eqn:methods_sr_pair_integer_definition} have been collected, the algorithm quests for a combination that results in a ratio between two square integers
\begin{equation}
    \begin{split}
        \prod_{r = 1}^{D} \tau_r \left(\dfrac{u_r - v_r \rsakey}{u_r}\right)
        &=
        \prod_{p_j \in \pbasis} {p_j}^{\sum_{r=1}^D \tau_r \tilde{e}_{j, r}}
        \\ &=
        {\bigg(\dfrac{X}{Y}\bigg)}^2.
    \end{split}
    \label{eqn:methods_sr_pairs_combination}
\end{equation}
Here, $\bm{\tau} \in \left\{0, 1\right\}^{D}$ is a binary vector to be determined. The product of $B$-smooth numbers on the left-hand side of Eq.~\eqref{eqn:methods_sr_pairs_combination} generates another $B$-smooth number. Given the property in Eq.~\eqref{eqn:methods_sr_pair_definition}, applying the modulo $\rsakey$ operation on both sides of Eq.~\eqref{eqn:methods_sr_pairs_combination} produces the desired congruence $X^2 \mathbin{/} Y^2 \equiv 1 \mod \rsakey$. To obtain a combination of smooth numbers with only even multiplicities, i.e., a square integer, we need to determine the binary vector $\bm{\tau}$. This is accomplished by solving the following system of linear equations~\cite{crandall2006_PrimeNumbers, boudot2022_review, galbraith2012_mathsOfpubkey}:
\begin{equation}
    \sum_{r=1}^D \tau_r \ \tilde{e}_{j, r} \equiv 0 \mod 2,
    \label{eqn:methods_linear_algebra_1}
\end{equation}
for $j = 0, \dots, \pi$. The system can be represented in matrix form as
\begin{equation}
    \mathcal{E} \cdot \bm{\tau} = 0,
    \label{eqn:methods_linear_algebra_2}
\end{equation}
transforming the problem into finding the kernel of a binary $\left(\pi + 1\right) \times D$ matrix $\mathcal{E}$, where the $r$-th column corresponds to the multiplicity vector $\bm{\tilde{e}}$ of the $r$-th collected sr-pair modulo $2$.
Theorems in linear algebra on the Galois field $\mathbb{GF}(2)$ of order $2$ ensure that a non-trivial solution to Eq.~\eqref{eqn:methods_linear_algebra_2} exists when the system has at least one free variable, i.e., when $D = \pi + 2$~\cite{bard2009_galois, galbraith2012_mathsOfpubkey}. This defines the minimum number of sr-pairs to be collected during the first stage of the algorithm. This post-collection processing step is a minor but common part of many integer factoring methods~\cite{lenstra1993_GNFS, pomerance2009_tale, pomerance2008_QuadraticSieve, rabah2006_reviewFIP, stevenhagen2008_GNFS} and efficient techniques are available to find the solution vector $\bm{\tau}$~\cite{coppersmith1993_linalgGF2, kleinjung2010_RSA-768record, boudot2020_RSA-250record}. It can be rigorously demonstrated~\cite{crandall2006_PrimeNumbers, pomerance2008_QuadraticSieve} that at least half of the solutions $\bm{\tau}$ to Eq.~\eqref{eqn:methods_linear_algebra_2} which yield the congruence $X^2 \equiv Y^2 \mod \rsakey$ also meet the non-triviality condition $X \not\equiv \pm Y \mod \rsakey$, thus leading to the factorization of $\rsakey$.

\textit{Collecting sr-pairs through lattice sieving.} Within this framework, Schnorr first proposed a lattice-based algorithm in Ref.~\cite{schnorr1991_diophantine}, where the task of collecting at least $D = \pi + 2$ sr-pairs is reframed as solving a series of optimization problems known as Closest Vector Problems (CVPs). A CVP entails determining the lattice point $\latticepoint^{op}$ in a given lattice $\lattice$ that is closest to a specified target point $\target$~\cite{galbraith2012_mathsOfpubkey}. Mathematically, we define the $\pi$-rank $(\pi + 1)$-dimensional lattice as the following infinite set of discrete points
\begin{equation}
    \lattice(\latticebasis)
    =
    \left\{
    \sum\nolimits_{j=1}^{\pi} e_j \latticepoint_j \ \big\vert \ \left\{e_j\right\}_{j=1}^{\pi} \in \mathbb{Z}^{\pi}
    \right\},
    \label{eqn:methods_lattice_definition}
\end{equation}
generated by the lattice basis
\begin{equation}
    \latticebasis = \left\{\latticepoint_1, \latticepoint_2, \dots, \latticepoint_{\pi}\right\}.
    \label{eqn:methods_lattice_basis_definition}
\end{equation}
Each $\latticepoint_j \in \mathbb{Z}^{\pi + 1}$ for $j = 1, 2, \dots, \pi$, and $\target \in \mathbb{Z}^{\pi + 1}$ is a target point not in the lattice $\lattice(\latticebasis)$. The problem asks for the lattice point $\latticepoint^{op} \in \lattice(\latticebasis)$ satisfying
\begin{equation}
    \lVert{\latticepoint^{op} - \target}\rVert
    =
    \min\limits_{\latticepoint \in \lattice(\latticebasis)} \lVert{\latticepoint - \target}\rVert,
    \label{eqn:methods_cvp_definition}
\end{equation}
in the Euclidean $\ell_2$-norm. Solving a generic instance of the CVP is NP-hard~\cite{emdeboas1981_cvpNPhardness, micciancio2001_hardnessCVP}. However, approximate techniques obtain nearly optimal solutions. For instance, Babai’s nearest plane algorithm~\cite{babai1986_BabaiAlgorithm, galbraith2012_mathsOfpubkey} solves this problem in polynomial time with the lattice rank $\pi$, provided Eq.~\eqref{eqn:methods_cvp_definition} is relaxed with an approximation factor $\eta = 2^{\order{\pi}}$. The algorithm finds an approximate closest vector $\latticepoint^{cl}$ such that
\begin{equation}
    \lVert{\latticepoint^{cl} - \target}\rVert
    \leq
    \eta \cdot \min\limits_{\latticepoint \in \lattice(\latticebasis)} \lVert{\latticepoint - \target}\rVert,
    \label{eqn:methods_approx_cvp_definition}
\end{equation}
within a number of operations that scales as shown in Ref~\cite{galbraith2012_mathsOfpubkey} with
\begin{equation}
    \order{
    \pi^6
    {\left[
    \log_{10}\left(\max\limits_{\latticepoint_j \in \latticebasis} \ \lVert{\latticepoint_j}\rVert\right)
    \right]}^3}.
    \label{eqn:methods_babai_complexity}
\end{equation}
Schnorr’s factoring algorithm~\cite{schnorr1991_diophantine, schnorr2021_classicalSchnorr, yan2022_quantumSchnorr} gets independent sr-pairs from approximate closest vector solutions to the $\rsakey$-related target
\begin{equation}
    \target_{\rsakey, \latticeparameter}
    =
    {\begin{pmatrix}
        0 & \cdots & 0 & \nint{10^\latticeparameter \ln{\rsakey}}
    \end{pmatrix}}^T,
    \label{eqn:methods_schnorr_targets}
\end{equation}
on a set of lattices described by the following $(\pi + 1) \times \pi$ lattice basis matrices:
\begin{align}
    \latticebasis_{\latticepermutation, \latticeparameter}
    =
    \begin{pmatrix}
        \latticepermutation(1) & 0 & \cdots & 0 \\
        0 & \latticepermutation(2) & \cdots & 0 \\
        \vdots & \vdots & \ddots & \vdots \\
        0 & 0 & \cdots & \latticepermutation(\pi) \\
        \nint{10^\latticeparameter \ln{p_1}} & \nint{10^\latticeparameter \ln{p_2}} & \cdots & \nint{10^\latticeparameter \ln{p_{\pi}}}
    \end{pmatrix},
    \label{eqn:methods_schnorr_lattices}
\end{align}
where $p_j$ for $j = 1, \dots, \pi$ are the prime numbers in the factoring basis $\pbasis$.
The columns of the lattice basis matrix in Eq.~\eqref{eqn:methods_schnorr_lattices} are built off the lattice basis vectors $\latticepoint_j \in \mathbb{Z}^{\pi + 1}$. A set of parameters can be varied to generate different CVPs. Specifically, $\latticeparameter$ is the so-called lattice precision parameter, while different diagonals of $\latticebasis$ are generated via a random permutation of the set $\left\{\nint{j \mathbin{/} 2}\right\}_{j=1}^{\pi}$, $\latticepermutation(j)$ being the $j$-th element in the permuted set. The symbol $\nint{\cdot}$ denotes the nearest integer rounding operation, with the convention that numbers with a fractional part of $0.5$ are rounded upwards.
In this lattice mapping of the factoring problem, each lattice point $\latticepoint \in \lattice(\latticebasis)$ corresponds uniquely to a pair of $p_{\pi}$-smooth integers $\left(u, v\right)$. Moreover, those lattice points closest to the target in each CVP $\singleCVP$ potentially offer sr-pairs. This stems from the above lattice construction, where sr-pairs represent approximate solutions to a system of simultaneous linear Diophantine equations for $\ln{\rsakey}$~\cite{schnorr1991_diophantine}. This factoring technique is known as \emph{Schnorr’s lattice sieving} because it filters sr-pairs that satisfy specific linear Diophantine equations for $\ln{\rsakey}$ using lattice-target pairs $\singleCVP$ associated with $\rsakey$.

\textit{Constructing sr-pairs from lattice points.} $B$-smooth pairs $\left(u, v\right)$  of positive integers are obtained from any lattice point $\latticepoint = \sum\nolimits_{j=1}^{\pi} e_j \latticepoint_j$ as follows:
\begin{equation}
    u = \prod_{\substack{e_j \geq 0 \\ j \in \left[1, \nqubits\right]}} p_j^{e_j};
    \qquad
    v = \prod_{\substack{e_j < 0 \\ j \in \left[1, \nqubits\right]}} p_j^{-e_j}.
    \label{eqn:methods_from_lattice_point_to_smooth_pair}
\end{equation}
Hence, the component vector $\bm{e} \in \mathbb{Z}^{\pi}$ of $\latticepoint$ on the lattice basis $\latticebasis$ translates to the multiplicity vectors of the integers $u$ and $v$. Given a lattice point, we compute these multiplicities by inverting the lattice basis matrix
\begin{equation}
    \bm{e} = \latticebasis^{-1} \cdot \latticepoint.
    \label{eqn:methods_lattice_point_components}
\end{equation}
This calculation is essential in Schnorr’s lattice sieving. The number of operations required for computing the matrix pseudo-inverse scales as $\order{\pi^3}$ for each lattice. Additionally, $\order{\pi^2}$ operations are needed for the matrix-vector multiplication in Eq.~\eqref{eqn:methods_lattice_point_components} each time we construct a smooth pair from a lattice point. From $\left(u, v\right)$, we form the integer $u - v\rsakey$ and decompose it into its prime factors to determine whether it is $B$-smooth, i.e., whether the lattice point $\latticepoint \sim \left(u - v\rsakey\right) \mathbin{/} u \sim \bm{\tilde{e}}$ yields an sr-pair for factoring. The time complexity for each $u - v\rsakey$ prime decomposition is bounded by a trial division strategy on the smoothness basis of size $\pi$, i.e., $\order{\pi}$, assuming finding remainder and division take place in constant time~\cite{crandall2006_PrimeNumbers}.

\textit{Schnorr’s sublinear theory.} Drawing from number theory and heuristic arguments, Schnorr proposed constructing CVPs with sublinear lattice dimensions~\cite{schnorr1991_diophantine, schnorr2021_classicalSchnorr}
\begin{equation}
    \pi = \dfrac{\bitlength}{\log_2\bitlength},
    \label{eqn:methods_sublinear_theory}
\end{equation}
alongside the corresponding sublinear smoothness bound for factoring $\rsakey$. Here $\bitlength = \lfloor{\log_2(\rsakey + 1)}\rfloor$ represents the size in bits of the input semiprime $\rsakey$ and $\lfloor{\cdot}\rfloor$ denotes the floor function. Schnorr asserted that this sublinear bound, combined with efficient lattice reduction methods to compute approximate CVP solutions, would break the security of RSA-2048 more efficiently than state-of-the-art classical methods~\cite{schnorr1991_diophantine, schnorr2021_classicalSchnorr}. However, this theoretical statement has sparked significant debate. Practical implementations of Schnorr’s sublinear classical algorithm have revealed discrepancies with his prediction~\cite{ducas2022_schnorr}. The primary shortcoming arises from the probability of a CVP approximate solution being associated with an sr-pair, which decays exponentially as the input bit-length increases.

\textit{Sublinear-resource quantum-classical Schnorr’s sieving.} To address the exponential complexity of the classical search for sr-pairs, B. Yan et al. have recently proposed in Ref.~\cite{yan2022_quantumSchnorr} a hybrid quantum-classical variant of Schnorr’s approach. This \emph{sublinear-resource quantum integer factorization} method combines classical Babai’s lattice reduction with quantum optimization, specifically the Quantum Approximate Optimization Algorithm (QAOA)~\cite{farhi2014_qaoa, blekos2024_qaoareview}. The goal is to extract more sr-pairs from each CVP by considering not only Babai’s solution $\latticepoint^{cl}$ but also a collection of “quantum” solutions that offer additional approximate closest lattice points to the target. The quantum approach builds on the following cost function
\begin{equation}
    F(x_1, \dots, x_{\pi})
    =
    {\left\lVert{
    \target - \latticepoint^{cl} - \sum_{j=1}^{\pi} \roundingop_j \lllvector_j}
    \right\rVert}^2,
    \label{eqn:methods_cvp_cost_function}
\end{equation}
where $\bm{\roundingop} = \left(\roundingop_1, \cdots, \roundingop_{\pi}\right) \in \left\{0, \pm 1\right\}^{\pi}$ are rounding optimization variables and $\lllbasis = \left\{\lllvector_1, \cdots, \lllvector_{\pi}\right\}$  is the LLL-reduced lattice basis of $\latticebasis$ obtained through the LLL algorithm~\cite{lenstra1982_LLL, galbraith2012_mathsOfpubkey}. Babai’s method is based on the LLL lattice reduction and outputs the lattice point $\latticepoint^{cl} = \sum\nolimits_{j=1}^{\pi} c_j \lllvector_j$, with $c_j = \nint{\mu_j}$ given by the nearest integer to the Gram-Schmidt coefficient obtained from the orthogonalization of $\latticebasis$~\cite{babai1986_BabaiAlgorithm}. Once the cost function in Eq.~\eqref{eqn:methods_cvp_cost_function} has been minimized, the rounding variables provide a more accurate or at least equivalent closest vector to the target:
\begin{equation}
    \latticepoint^{q}
    =
    \latticepoint^{cl} + \sum_{j=1}^{\pi} \roundingop_j \lllvector_j
    =
    \sum_{j=1}^{\pi} (c_j + \roundingop_j) \lllvector_j.
    \label{eqn:methods_cvp_quantum_solutions}
\end{equation}
To encode efficiently the possible values $-1$, $0$ and $+1$, i.e., all the possible rounding corrections to $\latticepoint^{bl}$ in Eq.~\eqref{eqn:methods_cvp_quantum_solutions}, each $\roundingop_j \in \left\{0, \pm 1\right\}$ is promoted to a quantum spin-1/2 operator, a.k.a. qubits,
\begin{equation}
    \roundingop_j \xrightarrow{} \hat{\roundingop}_j
    =
    \dfrac{1}{2}
    \mathrm{sign}\left(\mu_j - c_j\right)
    \left(\mathbb{1} - \hat{\sigma}_{j}^{z}\right),
    \label{eqn:methods_rounding_operators}
\end{equation}
where $ \hat{\sigma}_j^z $ denotes the $z$-Pauli matrix acting on the $j$-th qubit. We obtain the spin glass Hamiltonian encoding the optimization problem defined in Eq.~\eqref{eqn:methods_cvp_cost_function}
\begin{equation}
    \qham
    =
    \sum_{k=1}^{\pi + 1}
    {\left(
    t_k - b^{cl}_k - \sum_{j=1}^{\pi} d_{j, k} \roundingop_j
    \right)}^2,
    \label{eqn:methods_cvp_hamiltonian}
\end{equation}
which describes a system of $\nqubits = \pi$ all-to-all interacting qubits. For simplicity, we omit the quantum operator symbol whenever it is clear from the context. The Hamiltonian in Eq.~\eqref{eqn:methods_cvp_hamiltonian} takes in all potential roundings of the Gram-Schmidt coefficients within Babai’s nearest plane algorithm. In a scenario where computing the entire spectrum of an $\nqubits$-qubit system is feasible, all $2^\nqubits$ possible roundings around $\latticepoint^{cl}$ are available. These new lattice points can be used to construct smooth pairs and potentially collect more than one sr-pairs per CVP. Note that this approach involves accessing the excited states of $\qham$ for each CVP, rather than only the ground-state encoding the minimum of the cost function, as typically done in quantum approaches to classical optimization~\cite{lucas2014_isingNP}.

% --------------------------------------------------------------------------------
\subsection{Tensor network formulation}
\label{subsec:methods_ttn_formulation_details}
% --------------------------------------------------------------------------------

Tensor networks (TNs) are powerful variational Ans{\"a}tze tailored to efficiently represent quantum states taking advantage of the correlations and entanglement properties that characterize these states~\cite{schollwoeck2011_dmrgmps, orus2014_intrompspeps,montangero2018_book, silvi2019_anthology}.
Numerical methods rooted in TNs have demonstrated remarkable success in classically simulating the equilibrium and out-of-equilibrium behavior of quantum many-body systems, spanning from condensed matter to lattice gauge theories~\cite{wall2012_mpsoutofeq, vlaar2021_peps3D, tindall2024_tn4ibm, felser2020_ttnQLM, magnifico2021_qed}.
TN methods thus provide a representation of the exponentially large wave function of a quantum many-body system using the Schmidt decomposition to compress the entanglement content of the wave function. The efficiency of this decomposition is governed by the so-called bond dimension $\bonddim$, i.e., the Schmidt rank, which interpolates
between a product state with $\bonddim = 1$ and the exact, albeit unfeasible, state with $\bonddim = 2^{\nqubits\mathbin{/}2}$.

Among the various TN geometries available, our paper focuses on \emph{tree tensor networks} (TTNs). TTNs are among the most prominent geometries for investigating long-range interacting many-body systems, due to their hierarchical structure, improved connectivity, and computational efficiency regarding both bond dimension and system size \cite{silvi2019_anthology}. TTNs have recently shown promise in addressing hard classical optimization problems~\cite{cavinato2021_ttncancer}. These problems are typically associated with classical Hamiltonians that exhibit long-range two-body interactions among Ising spins~\cite{lucas2014_isingNP}. In the context of our research, we aim to leverage the TTN adaptability to enhance the demanding optimization task related to the search for sr-pairs within the spectrum of the CVP Hamiltonian $\qham$ defined in Eq.~\eqref{eqn:methods_cvp_hamiltonian}, which is a classical Hamiltonian, being diagonal within the chosen computational basis, and characterized by glassy all-to-all interactions.

Typically, the variational ground-state search~---~used as well in our approach~---~approximates the ground state of a physical system with a specific ansatz, e.g., a TTN state. As optimizations are local with respect to the tensor in the network and the bond dimension limits the entanglement, it usually takes many sweeps to converge to the ground state. Replicating this approach without modifications, we sample exactly one eigenstate of the system; the diagonal Hamiltonian has eigenstates with $\bonddim = 1$ which converges well. Established methods like excited state search do either scale unfavorably when requesting many excited states~\cite{jaschke2018_mps} or we have not explored them like the minimal entangled typical thermal state algorithm~\cite{stoudenmire2010_thermal}. Section~\ref{subsec:methods_ttn_sampling} explains the sampling method that we use for the tensor-network-enhanced Schnorr’s algorithm presented here; the algorithm fulfills the requirements in terms of the number of sampled states required for factoring.

The detailed scaling of the variational ground-state search is
\begin{equation}
    \order{\nqubits^3 \bonddim^4},
    \label{eqn:methods_ttn_gs_search_bigoh}
\end{equation}
where the contraction of an effective operator with a TTN tensor is the most costly step with $\order{\bonddim^4}$; one power of $\nqubits$ originates in the number of tensors to be updated, the other two powers from the bond dimension of the Hamiltonian representation. We ignore the number of sweeps due to the quick convergence.

% --------------------------------------------------------------------------------
\subsection{Sampling formalism}
\label{subsec:methods_ttn_sampling}
% --------------------------------------------------------------------------------

\begin{figure*}[t]
    \centering
    \begin{overpic}[width=0.85 \textwidth,unit=1mm]{ttn_sampling_performance_vs_bond_dim.png}
    \end{overpic}
    \caption{\emph{Relation between bond dimension and sampling smooth-relation pairs from a single lattice via tree tensor networks.}
    We plot both no-sr (gray points) and sr-pairs (green points) sampled from a lattice using the efficient TTN OPES method at different bond dimensions $\bonddim$ in the first row. The input RSA key has a bit-length of $\bitlength = 70$, and the CVP problem is characterized by $\nqubits = 32$. The number of samples is approximately $\bitlength^3 = 70^3$. The x-axis indicates the square root of the energy $\lVert\target - \latticepoint_j \rVert$ of the extracted eigenstate, i.e., the Euclidean distance to target $\target$ of the sampled pair identified with the lattice point $\latticepoint_j$. The y-axis represents the bit-length $\bitlength_j$ of the integer $u - v\rsakey$ associated with the sampled pair. Bond dimensions greater than $\bonddim = 8$ do not notably impact the sieving performance. Indeed, the average number of sr-pairs per lattice (AsrPL), i.e., the number of green points, is approximately $\asrpl \approx 400$ for each bond dimension.
    The second row shows the corresponding sampling probability for the first $400$ no-sr and sr-pairs ordered by decreasing probability. The x-axis denotes the pair’s numbering index $j$, while the y-axis shows the probability of the associated bit-string (gray bars for no-sr pairs, green bars for sr-pairs).}
    \label{fig:methods_ttn_sampling_with_probability}
\end{figure*}

The sampling formalism relies on three key aspects of the problem itself and tensor network methods: a)~the eigenstates are also eigenstates of the $z$-basis, i.e., any projective measurement in the $z$-basis yields a potential solution;
b)~an insufficiently converged TTN ground state overlaps with the excited states, although the probabilities are small;
c)~we can efficiently sample states in decaying probability tails with the algorithm presented in Ref.~\cite{ballarin2025_sampling}.
Thus, our goal is to launch a variational ground-state search which is not well-converged; the sampling yields then
excited states in the $z$-basis which are potentially low-energy smooth-relation pairs for Schnorr’s algorithm.

As the TTN converges well with the diagonal Hamiltonian $\qham$ even within one sweep, we introduce a perturbation breaking the diagonal form via random transverse fields $\left\{ \tfields(j) \right\}_{j=1}^{\nqubits}$
\begin{align}
    \qham^{\prime}
    =
    \qham + \sum_{j=1}^{\nqubits} \tfields(j) \, \sigma_{j}^{x} .
\end{align}
We use random couplings $\tfields(j)$ for the Pauli operator $\sigma_{j}^{x}$ acting on site $j$.
Specifically, the strengths of the local random transverse fields are controlled by a tunable hyper-parameter $\tfieldoom \in \mathbb{R}_{+}$, defined as
\begin{equation}
    \label{eqn:methods_transverse_field_strengths}
    \begin{split}
        &\tfields(j) \sim \mathscr{U} \left(- \left\vert\overline{\tfields}\right\vert, \left\vert\overline{\tfields}\right\vert\right) \\
        &\overline{\tfields} \sim \dfrac{1}{\tfieldoom}
        \langle \, \left\{W_{j j^\prime}\right\}, \, \left\{\mathrm{h}_j\right\} \rangle .
    \end{split}
\end{equation}
The parameter $\tfieldoom \in \mathbb{R}_{+}$ sets the relative magnitude of the transverse fields with respect to the diagonal spin-glass couplings $\left\{W_{j j^\prime}\right\}_{j < j^\prime = 1}^{\nqubits}$ and $\left\{\mathrm{h}_j\right\}_{j = 1}^{\nqubits}$~---~corresponding to the two-body spin interactions and single-spin bias terms in Eq.~\eqref{eqn:methods_cvp_hamiltonian}, respectively.
Here, $\mathscr{U}$ denotes the uniform distribution over the indicated interval, while $\langle{\circ}\rangle$ is the arithmetic mean.
By adjusting $\tfieldoom$, we control the scale of the off-diagonal (transverse) terms relative to the diagonal couplings.
Computationally, we define the special case $\tfieldoom = 0$ as the one corresponding to omitting the transverse-field term entirely, recovering the classical unperturbed CVP Hamiltonian in Eq.~\eqref{eqn:methods_cvp_hamiltonian}.
We carefully tune the amplitude of the random couplings $h_{x}(j)$ and then obtain TTN wave functions with sufficient overlap to excited states of the original Hamiltonian.

We use the sampling algorithm from Ref.~\cite{ballarin2025_sampling} out-of-the-box. Multiple super-iterations are carried out with exit criteria for the number of target states or the accumulated probability from sampled states. The ability to avoid re-sampling known states leads to an exponential speedup for sampling small
probabilities in exponentially decaying tails of distributions. We present the dataset of samples and their probability histogram at different bond dimensions $\bonddim$ in Figure~\ref{fig:methods_ttn_sampling_with_probability}, where Figure~\ref{fig:ttn_sampling_with_probability}a)-c) in Section~\ref{subsec:sampling_performance} contains the subset of this data for the $\bonddim = 8$ case.
Recall that the data are for our sieving procedure on a single lattice with $\bitlength = 70$ and $\nqubits = 32$. We plot the quality of the sampled pairs in terms of their distance from the target point ($x$-axis) and the size of the
resulting integer $u - v\rsakey$ in bits. We sample approximately $\bitlength^3$ states. We conclude that the impact of the bond dimension is minor here, while this parameter usually is key to convergence in simulations of entangled systems, e.g., condensed matter problems or lattice gauge theories. Note that a bond dimension $\bonddim = 8$ is sufficient to obtain an AsrPL of approximately $\asrpl = 400$ independent sr-pairs from a single lattice, even though their probabilities of being measured are on the order of $10^{-6}$.

The detailed scaling for the sampling including the ground-state search of Eq.~\eqref{eqn:methods_ttn_gs_search_bigoh} yields
\begin{align}
    \order{\nqubits^3 \bonddim^4 + \nqubits \bitlength^{\samplespower} \bonddim^4},
\end{align}
where we directly enter the number of samples as the bit-length $\bitlength$ and the meta parameter $\samplespower$. The sampling of the TTN state is independent of the bond dimension of the Hamiltonian: hence, the cost scales linearly with the number of qubits.

In the future, other techniques than the ground-state search can be exploited in combination with the sampling, e.g., finite-temperature representations~\cite{verstraete2004_mpsfinitetemp, zwolak2004_mixed1d, werner2016_tnopen}. The sampling of states can also extend beyond tensor network methods. Notably, this approach is also valuable for benchmarking quantum computer algorithms or quantum simulators devoted to Schnorr’s sieving. In particular, it is useful to understand the limitations and bottlenecks to be considered and overcome.

% --------------------------------------------------------------------------------
\subsection{Beyond Schnorr’s sublinear theory}
\label{subsec:methods_tnss_justification}
% --------------------------------------------------------------------------------

In this Section, we numerically demonstrate the potential advantage of the hybrid classical-quantum lattice sieving outlined in Ref.~\cite{yan2022_quantumSchnorr} over the original Schnorr’s factoring algorithm, provided that the sublinear scaling stated in Ref.~\cite{schnorr2021_classicalSchnorr}:
\begin{equation}
    \pi_1 = \pi_2 = \dfrac{\bitlength}{\log_2 \bitlength},
    \label{eqn:methods_sublinear_bounds}
\end{equation}
is relaxed. First, we implement Schnorr’s sublinear method to factorize RSA keys with increasing bit-lengths in the range $\bitlength \in \left[10, 60\right]$. Using the size defined in Eq.~\eqref{eqn:methods_sublinear_bounds} for both the factoring and the smoothness bases, the algorithm does not successfully factorize any of the considered bit-lengths. This is due to an insufficient number of sr-pairs found while sieving Babai’s solutions through sublinear dimensional lattices and evaluating their smoothness on $B = p_{\pi_2}$.

To increase the number of sr-pairs, we extend beyond the sublinear theory by setting the lattice rank $\pi_1$ to the theoretical sublinear value and gradually increase it by $1$ until a non-trivial factorization is obtained. The size of the smoothness basis is adjusted to $\pi_2 = 2 \pi_1^2$ as arbitrarily proposed in Ref.~\cite{yan2022_quantumSchnorr}, albeit without any justification.
In Figure~\ref{fig:methods_tnss_work_justification}, we plot the AsrPL and the number of CVPs necessary to compute the two prime factors, $p$ and $q$, of the RSA keys with bit-lengths up to $\bitlength = 60$. The values of the increased lattice rank $\pi_1$ used in these simulations are summarized in Table~\ref{table:methods_lattice_dimensions_justification}. The black squares indicate an exponentially vanishing $\asrpl$, resulting in an exponential number of CVPs required for factoring. This demonstrates that even relaxing the sublinear growth of $\pi_1$ and the smoothness bound $B$ with problem size, Schnorr’s lattice sieving requires an exponential number of resources.
\begin{table}[t]
    \centering
    \caption{\emph{Lattice ranks, i.e., the number of qubits, used to obtain the results in Figure~\ref{fig:methods_tnss_work_justification}}.
    When the sublinear theory defined in Eq.~\eqref{eqn:methods_sublinear_bounds} is not enough to successfully factorize an RSA key $\rsakey$ with bit-length $\bitlength$, we incrementally increase $\pi_1$ ($\nqubits$ in the classical-quantum Schnorr’s sieving) and set $\pi_2 = 2\pi_1^2$ until non-trivial prime factors $p$ and $q$ of $\rsakey$ are obtained.}
    \hfill
    \begin{ruledtabular}\begin{tabular}{c c c}
        \multicolumn{1}{p{2cm}}{\centering $\phantom{=}$ \\ Bit-length $\bitlength$ \\ $\phantom{=}$}
        &\multicolumn{1}{p{2cm}}{\centering Sublinear \\ lattice rank $\pi_1$}
        &\multicolumn{1}{p{2cm}}{\centering Increased \\ lattice rank $\pi_1$} \\
        \hline
        \rule{0pt}{3ex}10 & 3 & 4 \\
        \rule{0pt}{2ex}20 & 5 & 7 \\
        \rule{0pt}{2ex}30 & 6 & 8 \\
        \rule{0pt}{2ex}40 & 8 & 10 \\
        \rule{0pt}{2ex}50 & 9 & 12 \\
        \rule{0pt}{2ex}60 & 10 & 15 \\
    \end{tabular}\end{ruledtabular}
    \label{table:methods_lattice_dimensions_justification}
\end{table}

Following Ref.~\cite{yan2022_quantumSchnorr}, we reexamine Schnorr’s sublinear theory by implementing the spin glass mapping of the CVP set to factorize the same RSA keys. In this method, the lattice rank $\pi_1$ corresponds to the number of qubits $\nqubits$. Given the small values of $\nqubits = \bitlength \mathbin{/} \log_2 \bitlength$ when considering $\bitlength \in \left[10, 60\right]$, we exactly compute the $2^\nqubits$ spin glass configurations in the spectrum of each CVP Hamiltonian defined in Eq.~\eqref{eqn:methods_cvp_hamiltonian}, and thus obtain the AsrPL and the number of CVPs Hamiltonian as before. The blue diamonds in Figure~\ref{fig:methods_tnss_work_justification} show an improvement compared to the previous Schnorr’s factorizations. Despite achieving successful factorization with a sublinear number of qubits, the exponential decay (growth) of $\asrpl$ ($\ncvp$) persists. This issue has been investigated in other works, see Ref.~\cite{grebnev2023_pitfalls, khattar2023_commentGoogle, aboumrad2023_ionqSchnorr}.

Finally, we increase the sublinear bounds again following Table~\ref{table:methods_lattice_dimensions_justification} to provide a fairer comparison between the spin glass-based sieving and Schnorr’s algorithm. The red triangles in Figure~\ref{fig:methods_tnss_work_justification} suggest that by increasing the number of qubits beyond the sublinear scaling, we can shift the exponential decay of $\asrpl$ to larger bit-lengths. Indeed, both $\asrpl$ and $\ncvp$ appear constant in the plotted range of $\bitlength$. To obtain these data points, we access the full CVP spectra as before.

This preliminary investigation of the two lattice sieving methods prompted us to introduce the TNSS and use it to scale the simulations with $\nqubits$, thereby characterizing the computational resources needed to target RSA keys of significant size for modern cryptosystems.

\begin{figure*}[t]
    \centering
    \begin{overpic}[width=0.95 \textwidth,unit=1mm]{justification_schnorr_vs_quantum_exact.pdf}
    \end{overpic}
    \caption{\emph{Comparison between Schnorr’s classical sieving and the quantum-powered sieving.} We compare the resource requirement of different algorithms: the classical Schnorr’s algorithm (Babai)~\cite{schnorr2021_classicalSchnorr} with lattice size and smoothness bound beyond the original sublinear assumption (black squares); Yan et al.~\cite{yan2022_quantumSchnorr} implementation of the quantum version of Schnorr’s algorithm (blue diamonds) with sublinear scaling of the lattice size and smoothness bound; Yan et al.~\cite{yan2022_quantumSchnorr} implementation of the quantum version of Schnorr’s algorithm (red triangles) beyond sublinear scaling of the lattice and smoothness bound. The x-axes represent the bit-length of the input RSA number, while the y-axes carry the chosen relevant quantities to the considered algorithms, i.e. the number of CVPs $\ncvp$, left panel, and the average number of sr-pairs per lattice $\asrpl$, right panel. The implementation of the classical Schnorr’s algorithm (Babai) with sublinear resources does not appear in the plots as the factorization is never successful.}
    \label{fig:methods_tnss_work_justification}
\end{figure*}

% -----------------------------------------------------------------------------
\subsection{$\bm{\ttnrecord}$-bit factorization}
\label{subsec:methods_recorddetails}
% -----------------------------------------------------------------------------

Here, we report the details regarding the largest factorization of the $\ttnrecord$-bit RSA key
\begin{equation}
    \rsakey = \rsakeyrecord
    \label{eqn:methods_ttn_record_key}
\end{equation}
achieved using the TNSS algorithm. We sieve $\ncvp = \ncvprecord$ CVPs characterized by $\nqubits = \ttnnqubitsrecord$. We set $\samplespower = \ttngammarecord$ and find $\numsrpairsrecord$ sr-pairs on the smoothness bound $B = \sboundrecord$ using a basis of prime numbers of size $\pi_2 = \sbasisdimrecord$. By combining these sr-pairs, we obtain the following factorization:
\begin{equation}
    \begin{split}
        \rsakey
        =
        \ &\pfactorrecord \ \times \\ 
        \ &\qfactorrecord.
    \end{split}
    \label{eqn:methods_ttn_record_factorization}
\end{equation}
Notice that the number of sr-pairs needed to find a non-trivial solution is less than the theoretical value given by $\pi_2 + 1$. To avoid biases, the two factors were unknown to us before factorization, as we generated the key via the RSA number generator in~\cite{rsa_generator}.

We emphasize that the RSA number $\rsakey$ in Eq.~\eqref{eqn:methods_ttn_record_key} is a representative instance of a $\ttnrecord$-bit factorization problem, included for the purpose of reproducibility. Nonetheless, the TNSS algorithm successfully factorizes any $\ttnrecord$-bit RSA key, and its effectiveness is not limited to this specific semiprime example reported.

The computational time was on the order of days on a single Intel(R) Xeon(R) Gold 5320 CPU node.

% -----------------------------------------------------------------------------
\subsection{Analysis on the AsrPL scaling model}
\label{subsec:methods_chisquare}
% -----------------------------------------------------------------------------

The data points in Figure~\ref{fig:pairs_density_qubit_scaling}a) are obtained by rescaling the average number of sr-pairs per lattice (AsrPL) $\asrpl$ by the number of samples per CVP, $\bitlength^\gamma$, and the RSA input bit-length $\bitlength$ by an appropriate function of the number of qubits $\nqubits$. In this Section, we show how we selected the rescaling functions and how we determined the optimal rescaling parameter $\omega = \omega^\ast \approx \rescparam$.

The function AsrPL depends on both the number of qubits $\nqubits$ and the bit-length $\bitlength$. In the following, we analyze the average number of sr-pairs per lattice $\asrpl$ as a function of the number of qubits, rather than RSA bit-length. To this end, we test two alternative rescaling models:
\begin{equation}
    g_{\omega}(\bitlength) = {\bitlength}^\omega
\end{equation}
and
\begin{equation}
    g_{\xi}(\bitlength) = \frac{1}{\Gamma_2} \mathrm{e}^{\bitlength \mathbin{/} \xi}    
\end{equation}
with $\omega$, $\Gamma_2$ and $\xi$ as fit parameters.
As shown in Figure~\ref{fig:methods_rescaling_function_comparison}, these two models can be clearly distinguished within the range of our simulation data.

To identify the more appropriate rescaling model, we perform a statistical analysis of the rescaled data.
In Figure~\ref{fig:methods_rescaling_model_comparison}a), we plot the same simulation dataset for $\asrpl$ from Figure~\ref{fig:pairs_density_qubit_scaling}a), now as a function of the number of qubits $\nqubits$. Figure~\ref{fig:methods_rescaling_model_comparison}b) shows the data after rescaling by $g_{\omega}(\bitlength)$ on the x-axis and $\bitlength^\samplespower$ on the y-axis. The collapse onto a single curve persists across all instances explored, revealing a universal scaling behavior.
\begin{figure}[t]
    \centering
    \begin{overpic}[width=0.8 \columnwidth,unit=1mm]{Figures/g_omega_vs_g_xi_diff_extended}
    \end{overpic}
    \caption{\emph{Rescaling function comparison.}
    Comparison between two functions $g_{\omega}(\bitlength) = \bitlength^{\omega}$ (blue curve) and $g_{\xi}(\bitlength) = \frac{1}{\Gamma_2} \exp\left(\bitlength \mathbin{/} \xi\right)$ (orange curve), where $\omega = \omega^\ast = \rescparam$, $\xi = \xi^\ast = \bestxiexp$ and $\Gamma_2 \approx \bestgammaexp$, used to rescale the same data points of Figure~\ref{fig:pairs_density_qubit_scaling}a) but over the bit-length range. The optimal value of $\Gamma_2$ is computed through a least-squares procedure in $1 \mathbin{/} \Gamma_2$ to minimize the relative difference between the two functions, while optimal rescaling parameter values are derived from the bootstrapped chi-squared curves. The left $y$-axis is in log scale. The dashed grey curve shows the relative difference $\varepsilon_\mathrm{\scriptscriptstyle rel}(\bitlength) = \left(g_{\xi}(\bitlength) - g_{\omega}(\bitlength)\right) \mathbin{/} \left[\frac{1}{2} \left(g_{\omega}(\bitlength) + g_{\xi}(\bitlength)\right)\right]$ between the two functions.}
    \label{fig:methods_rescaling_function_comparison}
\end{figure}
\begin{figure*}[t]
    \begin{center}
        \begin{minipage}{0.95\textwidth}
            \begin{overpic}[width=1.0\columnwidth, unit=1mm]{effective_AsrPL_vs_effective_bitlength_poly}
                \put(1, 70){a)}
                \put(84, 70){b)}
            \end{overpic}
        \end{minipage}
    \end{center}
    \begin{center}
        \caption{
            \emph{Rescaling process.}
            \textbf{(a)}~The average number of sr-pairs per lattice (AsrPL) $\asrpl$ as a function of the number of qubits $\nqubits$. The different curves show the AsrPL for different values of the hyper-parameter $\gamma$ and RSA-key bit-length $\bitlength$. Each data point represents an average over $10$ random RSA keys. Each $\asrpl$ is computed using $\ncvp = 50$ lattices. The y-axis is plotted on a logarithmic scale. The red dashed line marks $\asrpl = 1$, corresponding to one useful pair per CVP on average.
            \textbf{(b)}~The same data as in panel \textbf{(a)}, but rescaled: the $\asrpl$ values are rescaled by $\bitlength^\gamma$ on the y-axis, and the number of qubits is rescaled by $g_{\omega^\ast}\left(\bitlength\right)$ on the x-axis, with $\omega^\ast \approx \rescparam$. Both the x-axis and y-axis are plotted on a logarithmic scale.
        }
        \label{fig:methods_rescaling_model_comparison}
    \end{center}
\end{figure*}

To quantify the statistical uncertainty and evaluate which model better fits the data, we apply a \emph{bootstrap} approach~\cite{chernick2008_bootstrap}. We fit the collapsed curves to a degree-$2$ polynomial for both rescaling schemes and perform $\bootstrapnsamples$ bootstrap re-samplings of the rescaled data (excluding simulation errors) to estimate the distribution of chi-squared values. Note that, since each data point is assigned equal uncertainty, the resulting chi-squared values represent unweighted sums of squared residuals.

This procedure is repeated across a range of $\omega$ and $\xi$ values near initial guesses to identify the minimum chi-squared value. The resulting Monte Carlo estimates of the average chi-squared and its statistical uncertainty are shown in Figure~\ref{fig:methods_BS_score_and_bootstrap}a). Our analysis demonstrates that the power-law rescaling consistently yields a lower (and thus better) chi-squared value than the exponential one.

To assess the confidence level in our model comparison, we compute an \emph{empirical p-value}. This quantifies how often the bootstrap procedure favors one rescaling over the other at their respective optimal parameters $\omega = \omega^\ast \approx \rescparam$ and $\xi = \xi^\ast \approx \bestxiexp$. Specifically, it provides a Monte Carlo estimate of probability $\mathrm{P}\left(\chi^2\left(\omega^\ast\right) < \chi^2\left(\xi^\ast\right)\right)$, that is, how often the power-law rescaling yields a lower chi-squared value than the exponential rescaling purely by chance~\cite{chernick2008_bootstrap}. The resulting p-value indicates a strong preference for the power-law model, with a value of approximately $0.9949$.

We further validate these findings using a second quantitative criterion to assess the quality of the data collapse under each rescaling model. Specifically, we employ the \emph{Bhattacharjee-Seno} (BS) collapse quality score, a well-established metric for evaluating how well numerical simulation datasets collapse onto a single universal curve under a proposed rescaling hypothesis~\cite{seno2001_collapsescore}.

The BS score, denoted by $S$, measures the average distance between each curve and the local average of all others in its neighborhood. The width of the local neighborhood is controlled by a tunable parameter $\epsilon \in \mathbb{R}$, which defines the width of the local comparison window.

Using the previously identified optimal rescaling parameters from the bootstrapped chi-squared test, we now compute the corresponding BS scores $S_{\omega^\ast}(\epsilon)$ and $S_{\xi^\ast}(\epsilon)$ over the range $\epsilon \in [0.05, 5]$. The results are shown in Figure~\ref{fig:methods_BS_score_and_bootstrap}b).

We find that the BS score is consistently lower for polynomial rescaling across the entire range of $\epsilon$, indicating a more coherent and tighter data collapse. This second independent statistical measure further supports our conclusion that the power-law rescaling ($g_{\omega}(\bitlength)$) provides a more accurate and robust description of the scaling behavior than the exponential alternative ($g_{\xi}(\bitlength)$).
\begin{figure*}[t]
    \centering
    \begin{overpic}[width=0.95 \textwidth,unit=1mm]{AsrPL_vs_nqubits_statistical_analysis}
        \put(2, 70){a)}
        \put(90, 70){b)}
    \end{overpic}
    \caption{
        \emph{Statistical analysis and model comparison under rescaling transformations.}
        \textbf{(a)} The bootstrapped chi-squared statistics~\cite{chernick2008_bootstrap} for evaluating the quality of the two rescaling models. For each model, the rescaled data are fitted using a degree-$2$ polynomial, and the corresponding chi-squared values are computed from $\bootstrapnsamples$ bootstrap samples across varying rescaling parameter values centered in an initial guess. The blue curve shows the result for the polynomial rescaling model, which exhibits a clear minimum at $\omega = \omega^{\ast} \approx \rescparam$ with a value of ${\chi}^2_{\omega^{\ast}} = \bestchipoly$, indicating a well-defined optimal rescaling. In contrast, the orange curve for the exponential rescaling model displays a relatively flat profile with a minimum at $\xi = \xi^{\ast} \approx \bestxiexp$ with $\chi^2_{\xi^{\ast}} = \bestchiexp$, suggesting a less robust optimal parameter. The empirical p-value derived from the comparison strongly favors the polynomial rescaling model, with a value of approximately $0.9949$.
        \textbf{(b)} The Bhattacharjee-Seno (BS) collapse quality score $S_{\scriptscriptstyle \mathrm{R}}\left(\epsilon\right)$~\cite{seno2001_collapsescore} is plotted as a function of the neighborhood width parameter $\epsilon \in \left[0.05, 5.0\right]$.
        The blue curve corresponds to the polynomial rescaling model using the function $g_{\omega}\left(\bitlength\right)$, while the orange curve corresponds to the exponential rescaling model with $g_{\xi}\left(\bitlength\right)$. Both curves are computed using the optimal rescaling parameters $\omega^{\ast} = \rescparam$ and $\xi^{\ast} = \bestxiexp$, respectively, determined via a bootstrap-based analysis of the chi-squared statistic (panel \textbf{(a)}). The y-axis is plotted on a logarithmic scale.
    }  
    \label{fig:methods_BS_score_and_bootstrap}
\end{figure*}
%
%%% ******* %%%
%%% ******* %%%

%%% References %%%
%%% ********** %%%
\clearpage
\bibliography{refs}
%%% ********** %%%
%%% ********** %%%

\end{document}